\documentclass[twocolumn, aps, pra, showpacs, superscriptaddress, floatfix, nofootinbib, 10pt]{revtex4-1}

\usepackage{amsmath,bbm,amsthm}
\usepackage{amssymb}
\usepackage{graphicx}
\usepackage{color}
\usepackage[colorinlistoftodos]{todonotes}
\usepackage[colorlinks=true, allcolors=blue]{hyperref}
\usepackage{dsfont}
\usepackage{float}
\usepackage{cancel}
\usepackage[autostyle]{csquotes}
\usepackage{enumitem}

\newtheorem{theorem}{Theorem}

\newtheorem{step}{Step}

\newtheorem*{theorem*}{Theorem}

\newcommand{\beq}{\begin{equation}}
\newcommand{\eeq}{\end{equation}}
\newcommand{\beqa}{\begin{eqnarray}}
\newcommand{\eeqa}{\end{eqnarray}}
\newcommand{\bra}[1]{\ensuremath{\left\langle#1\right|}}
\newcommand{\ket}[1]{\ensuremath{\left|#1\right\rangle}}

\def\half{\frac{1}{2}}
\def\opone{\leavevmode\hbox{\small1\normalsize\kern-.33em1}}

\newcommand{\dd}{\mathrm{d}}

\renewcommand{\today}{\number\day\space\ifcase\month\or
   January\or February\or March\or April\or May\or June\or
   July\or August\or September\or October\or November\or December\fi
   \space\number\year}

\begin{document}

\title{Genuine quantum nonlocality in the triangle network}
\date{\today}
\author{Marc-Olivier Renou}
\affiliation{D\'epartement de Physique Appliqu\'ee, Universit\'e de Gen\`eve, CH-1211 Gen\`eve, Switzerland}

\author{Elisa B\"aumer}
\affiliation{Institute for Theoretical Physics, ETH Zurich, Wolfgang-Pauli-Str. 27, 8093 Z\"urich, Switzerland} 

\author{Sadra Boreiri}
\affiliation{School of Computer and Communication Sciences, École Polytechnique Fédérale de Lausanne, CH-1015 Lausanne, Switzerland}

\author{Nicolas Brunner}
\affiliation{D\'epartement de Physique Appliqu\'ee, Universit\'e de Gen\`eve, CH-1211 Gen\`eve, Switzerland}

\author{Nicolas Gisin}
\affiliation{D\'epartement de Physique Appliqu\'ee, Universit\'e de Gen\`eve, CH-1211 Gen\`eve, Switzerland}

\author{Salman Beigi}
\affiliation{School of Mathematics, Institute for Research in Fundamental Sciences (IPM), Tehran,
Iran}

\begin{abstract}
Quantum networks allow in principle for completely novel forms of quantum correlations. In particular, quantum nonlocality can be demonstrated here without the need of having various input settings, but only by considering the joint statistics of fixed local measurement outputs. However, previous examples of this intriguing phenomenon all appear to stem directly from the usual form of quantum nonlocality, namely via the violation of a standard Bell inequality. Here we present novel examples of ``quantum nonlocality without inputs'', which we believe represent a new form of quantum nonlocality, genuine to networks. Our simplest examples, for the triangle network, involve both entangled states and joint entangled measurements. A generalization to any odd-cycle network is also presented. Finally, we conclude with some open questions.
\end{abstract}
\maketitle

\section{Introduction}\label{introduction}

Bell's theorem is arguably among the most important results in the foundations of quantum theory \cite{Bell}. It also had a major influence on the development of quantum information science \cite{Ekert}, and led recently to the so-called device-independent paradigm \cite{BHK,acin,colbeck,pironio}.

 \begin{figure}[b]
\centering
\includegraphics[width=0.8\columnwidth]{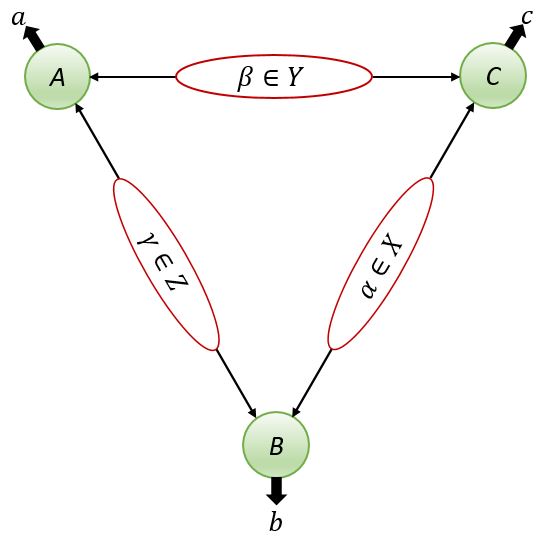}
\caption{The triangle network features three observers (green circles), connected by three independent bipartite sources (red ovals). Here the sources distribute local variables (i.e. shared randomness).}
\label{function}
\end{figure}

In his seminal work, Bell demonstrated that two distant observers, performing local measurements on a shared entangled state, can establish strong correlations which cannot be explained in any physical theory satisfying a natural principle of locality.
 These nonlocal quantum correlations can be demonstrated experimentally using Bell inequalities.
Recently, long-awaited loophole-free tests of quantum nonlocality were finally reported, providing the basis for the implementation of device-independent quantum information protocols \cite{hensen, shalm, giustina}. 

An interesting direction is to understand quantum nonlocality in scenarios involving more than two observers.
The standard approach to this problem (referred to as multipartite Bell nonlocality) considers three (or more) distant observers sharing an entangled state distributed by a common source, and leads to interesting new effects; see e.g. \cite{review} for a review.
This represents the simplest generalization of quantum nonlocality to the multipartite case, and most of the concepts and tools developed for bipartite nonlocality can generally be directly extended here.

Recently, a completely different approach to multipartite nonlocality was proposed \cite{branciard,branciard2,fritz}, focusing on quantum networks.
Here, distant observers share entanglement distributed by several sources which are assumed to be independent from each other.
By performing joint entangled measurements (such as the well-known Bell state measurement used in quantum teleportation \cite{bennett}), observers may correlate distant quantum systems and establish strong correlations across the entire network.
Typically, each source connects here only a strict subset of the observers.
It turns out that this situation is fundamentally different from standard multipartite nonlocality, and allows for radically novel phenomena.
As regards correlations, it is now possible to witness quantum nonlocality in experiments where all the observers perform a fixed measurement, i.e. they receive no input \cite{fritz,branciard2,fraser,gisin,NG2018,Renou}. This effect of quantum nonlocality without inputs is remarkable, and radically departs from previous forms of quantum nonlocality. 

So far, however, all known examples of quantum nonlocality without inputs can be traced back to standard Bell inequality violation. This naturally leads to the question of whether completely novel forms of quantum nonlocality, genuine to the network configuration, could arise. Here we address this question, by presenting an instance of quantum nonlocality in the triangle network, which we argue is fundamentally different from previously known forms of quantum nonlocality. In particular, our construction crucially relies on the combination of shared entangled states and joint entangled measurements performed by the observers. We present several generalizations of our main result, in particular to any cycle network featuring an odd number of parties. We conclude with a discussion and comment on the main open questions.

\section{Scenario and main result}\label{qubit_exple}

We consider the so-called triangle quantum network sketched in Fig. 1. It features three observers (Alice, Bob and Charlie). Every pair of observers is connected by a (bipartite) source, providing a shared physical system (represented e.g. by a classical variable or by a quantum state). Importantly, the three sources are assumed to be independent of each other. Hence, the three observers share no common (i.e. tripartite) piece of information. Based on the received physical resources, each observer provides an output ($a$, $b$ and $c$, respectively). Note that the observers receive no input in this setting, contrary to standard Bell nonlocality tests. The statistics of the experiment are thus given by the joint probability distribution $P(a,b,c)$.

Characterizing the set of distributions $P(a,b,c)$ that can be obtained from physical resources (in particular classical or quantum) is a highly non-trivial problem. The main difficulty stems from the assumption that the sources are independent. This makes the set of possible distributions $P(a,b,c)$ non-convex, and standard methods used in Bell nonlocality are thus completely unadapted to this problem. Strong bounds on the limits of classical correlations are thus still missing, which in turn renders the discussion of quantum nonlocality in the triangle network challenging.

Here we follow a different approach in order to present instances of quantum nonlocality in the triangle network.
Specifically, we first construct explicitly a family of quantum distributions $P_Q(a,b,c)$, using both entangled quantum states (distributed by each of the three sources), and entangled joint measurements performed by each observer.
Then we show that these quantum distributions cannot be reproduced by any ``trilocal'' model, i.e. a local model ``a la Bell'' where all three sources are assumed to be independent from each other. Formally, we prove that 
\begin{align} \label{trilocal}
P_Q(a,b,c) &\neq \\\int \dd\alpha \int& \dd\beta \int \dd \gamma     P_A(a|\beta, \gamma) \, P_B(b|\gamma,\alpha) \, P_C(c|\alpha,\beta)   \nonumber
\end{align}
where $\alpha\in X$, $\beta\in Y$ and $\gamma\in Z$ represent the three local variables distributed by each source and $P_A(a|\beta, \gamma),P_B(b|\gamma,\alpha),P_C(c|\alpha,\beta)$ represent arbitrary deterministic response functions for Alice, Bob and Charlie. Our proof does not rely on the violation of some Bell-type inequality, but is based on a logical contradiction. More precisely, we first identify a certain number of necessary properties that any trilocal model should have in order to reproduce $P_Q(a,b,c)$, and then show that these properties cannot all be satisfied at the same time.

Let us now construct explicitly our quantum distributions $P_Q(a,b,c)$. Each source produces the same pure maximally entangled state of two qubits, 
\begin{equation*}
\ket{\psi_\gamma}_{A_\gamma B_\gamma} = \ket{\psi_\alpha}_{B_\alpha C_\alpha} = \ket{\psi_\beta}_{C_\beta A_\beta}  = \frac{1}{\sqrt{2}} ( \ket{00} +  \ket{11}) \,.
\end{equation*}
Note that each party receives two independent qubit subsystems; for instance Alice receives subsystems $A_\beta$ and $A_\gamma$. Next, each party performs a projective quantum measurement in the same basis. In the following, we use the basis (a set depending on one real parameter $u$) given by
\begin{align} \label{BSM}
\ket{\uparrow}&=\ket{01} &  \ket{\chi_0}=u \ket{00} + v \ket{11}  \nonumber \\
\ket{\downarrow}&=\ket{10} &  \ket{\chi_1}=v \ket{00} -u \ket{11}
\end{align}
with $u^2+v^2=1$ and $0<v<u<1$.  For Alice, we label it $\{\ket{\phi_a}_{A_\beta A_\gamma}\}$ for $\phi_a \in\{\uparrow,\downarrow,\chi_0,\chi_1\}$ and adopt similar notations for Bob and Charlie.
Remark that only two out of the four states in that basis are entangled.
The statistics of the experiment are given by 
\begin{align*}
P_Q(a,b,c) = |\bra{\phi_a} \bra{\phi_b} \bra{\phi_c}\ket{\psi_\gamma} \ket{\psi_\alpha} \ket{\psi_\beta}|^2,
\end{align*}
where we did not specify the Hilbert spaces supporting the states. Note that when evaluating $P_Q(a,b,c)$, one should be attentive to which Hilbert space support each state and measurements. 

We now state the main result of this letter:
\begin{theorem}\label{theorem_qubit}
The quantum distribution $P_Q(a,b,c)$ cannot be reproduced by any classical trilocal model (in the sense of Eq. \eqref{trilocal}) when $u_{\mathrm{max}}^2< u^2<1$, where $u_{\mathrm{max}}^2 = \frac{-3+(9+6\sqrt{2})^{2/3}}{2(9+6\sqrt{3})^{1/3}} \approx 0.785$
\end{theorem}

We now sketch the proof; all details are given in Appendix A. The main idea is that the quantum distribution $P_Q(a,b,c)$ features a certain number of specific constraints. Indeed, one has that 
\begin{align}\label{constraint1}
P_Q(a=\uparrow,b=\uparrow) = P_Q(a=\downarrow,b=\downarrow)  =0
\end{align}
Symmetric relations are obtained by permuting the parties. Also, the number of parties that have an output in $\chi=\{\chi_0,\chi_1\}$ must be odd. Moreover, introducing the notation $u_0 = -v_1 =  u$ and $v_0 =  u_1 = v$ (such that $\ket{\chi_t}=u_t \ket{00} + v_t \ket{11}$) we have that 
\begin{align}
P_Q(\chi_i, \uparrow,\downarrow)  = \frac{1}{8}u_i^2,~~& 
P_Q(\chi_i, \downarrow,\uparrow)  = \frac{1}{8}v_i^2,\label{constraint2}\\
P_Q( \chi_i, \chi_j , \chi_k) = \frac{1}{8} (&u_i u_j u_k + v_i v_j v_k)^2 \label{constraint3}
\end{align}
and similar relations by permuting the parties. 
These four properties are essentially all we need. Indeed, for some specific choice of the measurement parameter $u$, no trilocal model can be compatible with all these four constraints at once. We prove by contradiction, assuming a trilocal model, in two successive steps where we identify conditions that this trilocal model reproducing $P_Q(a,b,c)$ should fulfill, to finally arrive at a contradiction. 
\begin{figure}
\centering
\includegraphics[width=0.9\columnwidth]{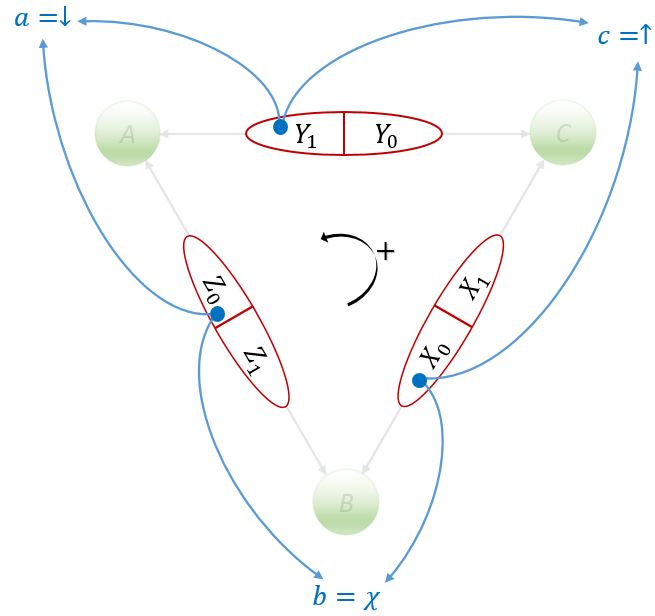}
\caption{Step~\ref{Step1} shows that any trilocal model compatible with the quantum distribution $P_Q(a,b,c)$ must have a specific structure. Specifically, for each source, the classical variable set can be divided into two equal weight (i.e. $P(\alpha \in X_0) = ... = 1/2$) disjoint subsets containing all output information on the coarse grained distribution where $\chi=\{\chi_0,\chi_1\}$ group together two outputs. For instance, when $\alpha \in X_0$, $\beta \in Y_1$ and $\gamma \in Z_0$, the outputs must be $a = \downarrow$ for Alice, $b = \chi$ for Bob, $c = \downarrow$ for Charlie. Note that the ordering is important.} \label{figure_step1}
\label{function}
\end{figure}

\begin{step}\label{Step1}
Here we consider the coarse graining of the output set $\{\uparrow,\downarrow,\chi=\{\chi_0,\chi_1\}\}$. We show that the sources sets can be partitioned in two subsets of equal weight $X=X_0\amalg X_1, Y=Y_0\amalg Y_1, Z=Z_0\amalg Z_1$ such that upon receiving $\beta$ and $\gamma$, Alice outputs
$(i)$ $a = \uparrow$ if she receives $\beta\in Y_0$ and $ \gamma \in Z_1$, 
$(ii)$ $a=\downarrow$ if she receives $\beta \in Y_1$ and $\gamma \in Z_0$,
$(iii)$ $a = \chi$ otherwise (similarly for Bob, Charlie, see Fig.~\ref{figure_step1}).
\end{step}
\begin{proof}
See Appendix \ref{mainresult}, it relies on \eqref{constraint1},\eqref{constraint2} and \eqref{constraint3}.
\end{proof} 
\begin{step}\label{Step2}
Let us introduce the following probability distribution 
\begin{align}
q(i,j&,k,t) := \label{q_ijkt} \\ &4 p(a=\chi_i,b=\chi_j,c=\chi_k, (\alpha,\beta,\gamma)\in (X_t,Y_t,Z_t)) \,.\nonumber
\end{align}
The following marginal distributions of $q(i,j,k,t)$ satisfy:
\begin{align}
q(i,j,k)&= \sum_t q(i,j,k,t)  = \frac{1}{2}( u_i u_j u_k +  v_i v_j v_k )^2, \label{marg_q_ijk}\\
q(i,t=0)&= \sum_{j,k} q(i,j,k,t=0)  =\frac{1}{2} u_i^2,\label{marg_q_it}
\end{align}
and similar constraints on $q(i,t=1)$, $q(j,t)$ and $q(k,t)$.
\end{step}
\begin{proof}
From Step~1, one can see that all parties output $\chi$ iff $(\alpha,\beta,\gamma)\in (X_t,Y_t,Z_t)$ with $t=0$ or $t=1$. This ensures that $q(i,j,k,t) $ is properly normalized. 
\eqref{marg_q_ijk} is straightforward from \eqref{constraint3}. \eqref{marg_q_it} can be deduced from Step 1 and the fact that Alice's output must be independent of $\alpha$ (see App. \ref{mainresult} for a more detailed proof). 
\end{proof}

At this point we arrive at a contradiction. Indeed, if a trilocal model existed, one should be able to define a distribution $q(i,j,k,t) $ that is compatible with all its marginals, in particular those marginals discussed above. However, this is not possible for all values of the parameter $u$ (which quantifies the degree of entanglement of measurement $\chi_0,\chi_1$), specifically when $0,785\approx u_{\mathrm{max}}^2<  u^2<1$. This concludes the proof.

A natural question is whether the distribution $P_Q$ is trilocal when $u^2  \leq u_{\mathrm{max}}^2 $. In Appendix~D we show that this is the case, by constructing an explicit trilocal model for $u^2   = u_{\mathrm{max}}^2$ (up to machine precision). We conjecture that $P_Q$ remains trilocal up to $u^2< u_{\mathrm{max}}^2$. Note that this can be proven for the case $u^2=1/2$. Here the trilocal model is obtained from Step~\ref{Step1}, with $\chi$ replaced by a uniformly random choice between $\chi_0$ and $\chi_1$.

Before entering a more general discussion about the implications of Theorem 1 and some natural open questions, we now briefly present several generalizations of the result.

\section{Generalisations}
The first extension considers the same scenario as in Theorem 1, with the difference that all sources now produce the same general entangled two-qubit pure states $\lambda_0 \ket{00} + \lambda_1 \ket{11}$ where $\lambda_0^2+\lambda_1^2=1$. We consider the same measurements \eqref{BSM}. In this case, Theorem 1 can be extended, with the condition that $  u_{\mathrm{max}} (\lambda_0)\lesssim u<1$ (see Appendix A for details). Interestingly, the lower bound $u_{\mathrm{max}} (\lambda_0)$ takes its lowest value for non-maximally entangled states ($\lambda_0 = \sqrt{2/3}$). In this case, we find $u_{\mathrm{max}} (\lambda_0) = \sqrt{2/3}$, implying that the projective joint measurement must feature non-maximally entangled states.

A second generalization considers the triangle network with higher dimensional quantum systems. Specifically, all three sources now produce a maximally entangled two-qutrit state, i.e. $\ket{\phi_3}=(\ket{00}+\ket{11}+\ket{22})/\sqrt{3}$. Each party performs the same joint entangled projective measurement, with nine outcomes, labeled by $a \in \{ 
\tilde{0}, \tilde{1}, \tilde{2}, \chi_0^\uparrow, \chi_1^\uparrow, \chi_2^\uparrow, \chi_{0}^\downarrow, \chi_{1}^\downarrow, \chi_{2}^\downarrow\}$ for Alice. The projectors are on the states
\begin{align*}
&\ket{\tilde{0}}:=\ket{00},~~  \ket{\tilde{1}}:=\ket{11},~~  \ket{\tilde{2}}:=\ket{22},\\
&\ket{\chi_i^\uparrow}=\eta_i^{01}\ket{01}+\eta_i^{02}\ket{02}+\eta_i^{12}\ket{12}, \\
&\ket{\chi_{i}^\downarrow}=\eta_i^{10}\ket{10}+\eta_i^{20}\ket{20}+\eta_i^{21}\ket{21}. 
\end{align*}
where the coefficients $\{\eta_i^{01},\eta_i^{02},\eta_i^{12}\}$ and $\{\eta_i^{10},\eta_i^{20},\eta_i^{21}\}$ are real and chosen such that the nine vectors form an orthonormal basis. Similarly to Theorem 1, one can show that for an appropriate choice of these parameters, the resulting quantum distribution is incompatible with any trilocal model (see Appendix B).

A third generalization explores networks beyond the triangle. 
Specifically, we prove a generalization of Theorem 1 for any $N-$cycle network, with $N$ being odd. 
Here all $N$ sources produce a maximally entangled two-qubit state, and all parties perform the same joint measurement, as in Eq. \eqref{BSM}. 
We show that for any $N$, the quantum distribution is incompatible with any $N$-local model (i.e. a straightforward generalization from Eq. \eqref{trilocal}) when the measurement parameter $u$ goes asymptotically to 1. 
Our approach cannot be directly adapted to even cycles.

\section{Discussion}
%
We presented novel examples of quantum nonlocality without inputs, mainly for the triangle network.
We believe that these examples represent a form of quantum nonlocality that is genuine to the network configuration, in the sense that it is not a consequence of standard forms of Bell nonlocality. 
These examples fundamentally differ from the one presented by Fritz in \cite{fritz}, relying on the violation of a standard bipartite Bell inequality. Let us first briefly review it.

Fritz's example can be viewed as a standard Bell test, embedded in the triangle network.
Consider that Alice and Bob share a two-qubit Bell state, with the goal of violating the CHSH Bell inequality.
Testing the CHSH inequality requires of course local inputs for both Alice and Bob. Although the triangle network features no explicit inputs, here effective inputs are provided by the two additional sources: the source connecting Alice and Charlie (resp. Bob and Charlie) provides a shared uniformly random bit, which is used as Alice's (resp. Bob's) input for the CHSH test.
All parties output the  ``input bits'' he receives. The correspondence between these outputs ensures that Alice's (resp. Bob's) output only depends on the source she (resp. he) shares with Charlie.
Finally, Alice and Bob both additionally output the output of their local measurement performed on the shared Bell state.
If this quantum distribution could be reproduced by a trilocal model, it would follow that local correlations can violate the CHSH inequality, which is impossible.

Let us comment on some significant differences between Fritz's construction and our example of Theorem 1. First, our construction has a high level of symmetry (all sources distribute the same entangled state and all measurements are the same) with only four outputs per party. In particular, it involves an entangled state for each source, whereas the example of Fritz requires entanglement for only one source (it can be symmetrized, but at the cost of adding new outputs). Moreover, our construction appears to rely on the use of joint measurements with entangled eigenstates, while Fritz's model uses only separable measurements. Hence Fritz's construction could be obtained from PR-boxes \cite{PR}. As the equivalent of joint measurements does not exist for PR-boxes \cite{barrett,short}, we believe that our example cannot be obtained from PR-boxes. 

Note that all the above arguments are only based on qualitative and intuitive arguments. We have no formal proof that in order to obtain the distribution $P_Q(a,b,c)$ one actually requires all states to be entangled and/or joint entangled measurements. In fact, even formalizing the problem is difficult, any progress in this direction would be interesting. An idea would be to use the notion of ``self-testing'' \cite{MY}, for instance by proving that all shared quantum states must be two-qubit Bell states and/or that all local measurements must feature specific entangled eigenstates \cite{Renou_BSM,Bancal_BSM}. 

Another important aspect of our construction that must be discussed is noise tolerance. As such, Theorem 1 clearly applies only to the exact quantum distribution $P_Q(a,b,c)$, i.e.~in the noiseless case. The trilocal set being topologically closed, it is clear that $P_Q(a,b,c)$ must have a certain (possibly very weak) robustness to noise: when adding a sufficiently small amount of local noise to $P_Q(a,b,c)$, one should still obtain a quantum distribution that is incompatible with any trilocal model. 
A promising method would be to consider the qutrit example, the proof of which involves the Finner inequality that allows in principle for the presence of noise. However we did not succeed in obtaining reasonable noise tolerance of our result so far. Other methods could also help, such as the ``inflation'' technique \cite{wolfe} \footnote{The inflation method consists of a sequence of tests of rapidly increasing computational complexity. For the first tests in the sequence, implementable on a computer, it appears that even the nonlocality of $P_Q(a,b,c)$ cannot be detected. Nevertheless, as inflation is known to converge in the limit \cite{navascues}, this is still an interesting possibility to explore.}. This could provide a nonlinear Bell inequality violated by our example.

The possibility of generating randomness from quantum nonlocality without inputs is a further interesting question. In particular, it seems very likely that our quantum distribution $P_Q(a,b,c)$ contains some level of intrinsic randomness. It would be interesting to see how this randomness could be quantified in a device-independent manner (still assuming independence of the sources).

\emph{Acknowledgements.---}We thank Alex Pozas and Elie Wolfe for discussions. We acknowledge financial support from the Swiss national science foundation (Starting grant DIAQ, NCCR-QSIT).

\newpage
\onecolumngrid
\appendix

\setcounter{theorem}{0}
\setcounter{step}{0}

\section{Proof of main result} \label{mainresult}

In this section, we provide a complete proof of Theorem 1. We directly prove the generalized result where the shared sources are non maximally entangled states: Alice, Bob and Charlie now share general two-qubit pure entangled states $ \ket{\psi} = \lambda_0 \ket{00} + \lambda_1 \ket{11}$ where $\lambda_0,\lambda_1$ are real and $\lambda_0^2+\lambda_1^2=1$. The measurement performed by the parties are as in the main text, namely projective in the basis
\begin{align} 
\ket{\uparrow}&=\ket{01} &  \ket{\chi_0}=u_0 \ket{00} + v_0 \ket{11}  \nonumber \\
\ket{\downarrow}&=\ket{10} &  \ket{\chi_1}=u_1 \ket{00} + v_1 \ket{11} \label{BSM_appendix}
\end{align}
with real number $u=u_0=-v_1$, $v=u_1=v_0$, $u^2+v^2=1$. For symmetry reason, we assume w.l.o.g. that $0<v\leq u<1$.

The global quantum state shared by the parties is given by
\begin{align*}
\ket{\psi_{ABC}}&=\lambda_0^3 \ket{000000} + \lambda_1^3 \ket{111111}+ \lambda_0^2 \lambda_1 \left( \ket{100001}+\ket{000110}+\ket{011000} \right) + \lambda_0 \lambda_1^2 \left( \ket{100111} + \ket{111001} + \ket{011110} \right)\\
&=\lambda_0^3 \left(u \ket{\chi_0} + v \ket{\chi_1} \right)^{\otimes 3} + \lambda_1^3 \left( v\ket{\chi_0} - u \ket{\chi_1}\right)^{\otimes 3} +\lambda_0^2 \lambda_1 \big( u (\ket{\uparrow\downarrow\chi_0}+\ket{\downarrow\chi_0\uparrow}+\ket{\chi_0\uparrow\downarrow}) + v(\ket{\uparrow\downarrow\chi_1}+\ket{\downarrow\chi_1\uparrow} \\
 &\quad+\ket{\chi_1\uparrow\downarrow}) \big)+\lambda_0 \lambda_1^2 \big( v (\ket{\downarrow\uparrow\chi_0} + \ket{\uparrow\chi_0\downarrow} + \ket{\chi_0\downarrow\uparrow})  -u (\ket{\downarrow\uparrow\chi_1}+\ket{\uparrow\chi_1\downarrow}+\ket{\chi_1\downarrow\uparrow}) \big).
\end{align*}
From this last equation, we can determine a certain number of crucial properties of the distribution $P_Q(a,b,c)$. First, note that the number of parties that output $\chi$ (i.e. either $\chi_0$ or $\chi_1$) must be odd. Second, we observe that 
\begin{align}\label{P_QUpUp}
P_Q(a=\uparrow,b=\uparrow) = P_Q(a=\downarrow,b=\downarrow)  =0.
\end{align}
Moreover, 
\begin{align}\label{P_QUpDown}
P_Q(a=\uparrow, b=\downarrow)=P_Q(b=\uparrow, c=\downarrow)=\lambda_0^4 \lambda_1^2  \quad \textrm{and} \quad
P_Q(c=\downarrow,a=\uparrow)=\lambda_0^2\lambda_1^4.
\end{align}
We have other similar relations by permuting the parties. Finally, we have that 
\begin{align} 
P_Q(\chi_i,\uparrow,\downarrow)=\lambda_0^4 \lambda_1^2 u_i^2,&~~P_Q(\chi_i,\downarrow,\uparrow)=\lambda_0^2 \lambda_1^4 v_i^2\label{PQi}\\
P_Q( \chi_i, \chi_j , \chi_k)& = (\lambda_0^3 u_i u_j u_k + \lambda_1^3 v_i v_j v_k )^2.\label{PQijk}
\end{align}


We can now prove the following generalization of Theorem 1:
\begin{theorem}\label{theorem_qubit_appendix}
There exists no trilocal model reproducing $P_Q(a,b,c)$, whenever  $ u_{\mathrm{max}}^2(\lambda_0) < u^2 < 1$ (see Fig.~\ref{function}), where $u_{\mathrm{max}}(\lambda_0)$ is implicitly defined by the constraint:
\begin{align}
3(\lambda_0^3 u^2 v - \lambda_1^3 u v^2)^2- 3u^2 (\lambda_0^6+\lambda_1^6) + 2(\lambda_1^6 + (\lambda_0^3 u^3+\lambda_1^3 v^3)^2) + \lambda_0^6 +(\lambda_0^3 v^3-\lambda_1^3 u^3)^2 \geq 0.\label{ineq}
\end{align}
\end{theorem}
The proof of Theorem~\ref{theorem_qubit_appendix} follows from two steps. We first assume, by contradiction, that  a classical 3-local strategy exists.
In Step~\ref{step:appendix_qubit_triangle_1}, we consider the behavior for which the two outputs $\chi_0, \chi_1$ are grouped into a single output $\chi$. We show that in that case, one only needs one bit of information about each of the local hidden variables to find the outputs.
Using this restriction and exploiting relations \eqref{PQi}, \eqref{PQijk}, Step~\ref{step:appendix_qubit_triangle_2} shows that those restrictions can be exploited to compute marginals of a probability distribution grouping outputs and hidden variables. For a good choice of measurement basis, those marginals are incompatible.

\begin{step}\label{step:appendix_qubit_triangle_1}
We consider the coarse graining of the output set $\{\uparrow,\downarrow,\chi=\{\chi_0,\chi_1\}\}$. The sources sets can be partitioned in two subsets
\begin{equation}
X=X_0\amalg X_1,~~~~~~~~Y=Y_0\amalg Y_1,~~~~~~~~Z=Z_0\amalg Z_1,
\end{equation}
(with $P(X_0)=P(Y_0)=P(Z_0)=\lambda_0^2$, $P(X_1)=P(Y_1)=P(Z_1)=\lambda_1^2$), such that the sets from which the local variables $ \alpha$, $\beta$ and $\gamma$ are taken determine all outputs.
More precisely, Alice answers 
\begin{enumerate}[label=(\roman*)]
\item $a = \uparrow$ iff she receives $\beta\in Y_0$ and $ \gamma \in Z_1$, 
\item $a=\downarrow$ iff she receives $\beta \in Y_1$ and $\gamma \in Z_0$,
\item $a = \chi$ otherwise,
\end{enumerate}
and similarly for Bob and Charlie (with a direct orientation of the cycle, see Fig.~\ref{figure_step1}).
\end{step}

\begin{proof}
Let us define the sets
\begin{align*}
X_0&=\{ \alpha \ | \ \exists \gamma \ s.t. \ b(\gamma,\alpha)=\downarrow\}, \quad
X_1=\{ \alpha \ | \ \exists \gamma \ s.t. \ b(\gamma,\alpha)=\uparrow\}, \nonumber\\
X_0'&=\{ \alpha \ | \ \exists \beta \ s.t. \ c(\alpha,\beta)=\uparrow\}, \quad
X_1'=\{ \alpha \ | \ \exists \beta \ s.t. \ c(\alpha,\beta)=\downarrow\}, \nonumber\\
Y_0&=\{ \beta \ | \ \exists \alpha \ s.t. \ c(\alpha,\beta)=\downarrow\}, \quad
Y_1=\{ \beta \ | \ \exists \alpha \ s.t. \ c(\alpha,\beta)=\uparrow\}, \nonumber\\
Y_0'&=\{ \beta \ | \ \exists \gamma \ s.t. \ a(\beta, \gamma)=\uparrow\}, \quad
Y_1'=\{ \beta \ | \ \exists \gamma \ s.t. \ a(\beta, \gamma)=\downarrow\}, \nonumber\\
Z_0&=\{ \gamma \ | \ \exists \beta \ s.t. \ a(\beta,\gamma)=\downarrow\}, \quad
Z_1=\{ \gamma \ | \ \exists \beta \ s.t. \ a(\beta,\gamma)=\uparrow\}, \nonumber\\
Z_0'&=\{ \gamma \ | \ \exists \alpha \ s.t. \ b(\gamma,\alpha)=\uparrow\}, \quad
Z_1'=\{ \gamma \ | \ \exists \alpha \ s.t. \ b(\gamma,\alpha)=\downarrow\}.
\end{align*}
Note that \eqref{P_QUpUp} directly implies that:
\begin{equation}\label{disjoint}
X_0\cap X_1'=X_0'\cap X_1=\o,~~~~~~~~ Y_0\cap Y_1'=Y_0'\cap Y_1=\o,~~~~~~~~Z_0\cap Z_1'=Z_0'\cap Z_1=\o.
\end{equation}
We also have $X=X_0'\cup X_1$.
By contradiction, assume $\alpha^*\notin X_0'\cup X_1$.
Then, for all $\beta,\gamma$, $b(\gamma,\alpha^*)\neq\uparrow$ and $c(\alpha^*,\beta)\neq\uparrow$. 
As the number of $\chi$ is odd, with \eqref{P_QUpUp}, we deduce that for all $\beta,\gamma$, $a(\beta,\gamma)\neq\downarrow$ which is absurd, as Alice answers $\downarrow$ sometimes.
With similar proofs, considering \eqref{disjoint}, we obtain:
\begin{equation}\label{partition}
X_0\amalg X_1'=X_0'\amalg X_1=X,~~~~~~~~ Y_0\amalg Y_1'=Y_0'\amalg Y_1=Y,~~~~~~~~ Z_0\amalg Z_1'=Z_0'\amalg Z_1=Z.
\end{equation}

Next, we show that $\o=X_0\cap X_1$. The proof goes also by contradiction. Assume that $\alpha^*\in X_0\cap X_1$: there exist $\gamma_1,\gamma_2$ such that $b(\gamma_1,\alpha^*)=\downarrow$ and $b(\gamma_2,\alpha^*)=\uparrow$. Take any $\beta\in Y$.
As $b(\gamma_1,\alpha^*)=\downarrow$, by \eqref{P_QUpUp}, we have $c(\alpha^*,\beta)\neq \downarrow$. 
Similarly, as $b(\gamma_2,\alpha^*)=\uparrow$ we have $c(\alpha^*,\beta)\neq \uparrow$. Hence $c(\alpha^*,\beta)=\chi$. As the number of $\chi$ is odd, we can conclude with \eqref{P_QUpUp} that $a(\beta,\gamma_1)=\uparrow$ and $a(\beta,\gamma_2)=\downarrow$. 
Consider now any $\alpha\in X$. 
As $a(\beta,\gamma_1)=\uparrow$, with \eqref{P_QUpUp}, we have $c(\alpha,\beta)\neq \uparrow$. As $a(\beta,\gamma_2)=\downarrow$, with \eqref{P_QUpUp}, we have $c(\alpha,\beta)\neq \downarrow$. Hence $c(\alpha,\beta)=\chi$. Remember this is valid for any $\alpha,\beta$. As in the setup Charlie does not answer $\chi$ all the time, this is a contradiction, which finishes the proof. With similar proofs and considering \eqref{partition}, we obtain:
\begin{equation}
X_0= X_0',~~~~X_1'= X_1,~~~~Y_0= Y_0',~~~~Y_1'= Y_1,~~~~ Z_0= Z_0',~~~~Z_1'= Z_1.
\end{equation}
Hence:
\begin{align*}
&a(\beta,\gamma)=\uparrow~\mathrm{iff}~\beta\in Y_0=Y_0',\gamma\in Z_1=Z_1',
&a(\beta,\gamma)&=\downarrow~\mathrm{iff}~\beta\in Y_1=Y_1',\gamma\in Z_1=Z_1',\\
&b(\gamma,\alpha)=\uparrow~\mathrm{iff}~\gamma\in Z_0=Z_0',\alpha\in X_1=X_1', 
&b(\gamma,\alpha)&=\downarrow~\mathrm{iff}~\gamma\in Z_1=Z_1',\alpha\in X_0=X_0',\\
&c(\alpha,\beta)=\uparrow~\mathrm{iff}~\alpha\in X_0=X_0',\beta\in Y_1=Y_1',
&c(\alpha,\beta)&=\downarrow~\mathrm{iff}~\alpha\in X_1=X_1',\beta\in Y_0=Y_0'.\\
\end{align*}

It remains now to compute the probabilities of the different subsets. 
Note that:
\begin{align*}
P(a=\uparrow, b=\downarrow) &= P(X_0)P(Y_0)P(Z_1) \\
P(c=\downarrow, a=\uparrow) &= P(X_1)P(Y_0)P(Z_1)\\
P(b=\uparrow, c=\downarrow) &= P(X_1)P(Y_0)P(Z_0)
\end{align*}
Using \eqref{P_QUpDown} we can conclude that $P(X_0)=P(Y_0)=P(Z_0)=\lambda_0^2$ and $P(X_1)=P(Y_1)=P(Z_1)=\lambda_1^2$, which finishes the proof of Step~\ref{step:appendix_qubit_triangle_1}.
\end{proof}

\begin{step}\label{step:appendix_qubit_triangle_2}
Let us introduce
\begin{align}
q(i,j,k,t)& :=p\big(a=\chi_i,b=\chi_j,c=\chi_k,(\alpha,\beta,\gamma)\in X_t \times Y_t \times Z_t | (\alpha,\beta,\gamma)\in X_0 \times Y_0 \times Z_0 \cup X_1 \times Y_1 \times Z_1\big )\nonumber\\
&= \frac{1}{\lambda_0^6+\lambda_1^6} p(a=\chi_i,b=\chi_j,c=\chi_k,(\alpha,\beta,\gamma)\in X_t \times Y_t \times Z_t). \label{q_ijkt_appendix}
\end{align}
$q(i,j,k,t)$ is a probability distribution. Moreover, the following marginal distributions of $q(i,j,k,t)$ satisfy:
\begin{align}
q(i,j,k)&= \sum_t q(i,j,k,t) = \frac{(\lambda_0^3 u_i u_j u_k + \lambda_1^3 v_i v_j v_k )^2}{\lambda_0^6+\lambda_1^6} \label{app:marg_q_ijk_qubit}\\
q(i,t=0)=&\frac{\lambda_0^6}{\lambda_0^6+\lambda_1^6} u_i^2~~~~ 
q(i,t=1)=\frac{\lambda_1^6}{\lambda_0^6+\lambda_1^6} v_i^2\label{app:marg_q_it_qubit}\\
q(j,t=0)=&\frac{\lambda_0^6}{\lambda_0^6+\lambda_1^6} u_j^2~~~~ 
q(j,t=1)=\frac{\lambda_1^6}{\lambda_0^6+\lambda_1^6} v_j^2\label{app:marg_q_jt_qubit}\\
q(k,t=0)=&\frac{\lambda_0^6}{\lambda_0^6+\lambda_1^6} u_k^2~~~~ 
q(k,t=1)=\frac{\lambda_1^6}{\lambda_0^6+\lambda_1^6} v_k^2\label{app:marg_q_kt_qubit}
\end{align}
\end{step}

\begin{proof}
The parties all answer $\chi$ iff $(\alpha,\beta,\gamma)$ are either in $X_0 \times Y_0 \times Z_0$ or in $X_1 \times Y_1 \times Z_1$. Hence $q(i,j,k,t)$ is a probability distribution.

\eqref{app:marg_q_ijk_qubit} follows directly from Eq. \eqref{PQijk}. As $q(i,j,k,t)$ is a probability distribution, \eqref{app:marg_q_it_qubit} can be deduced from \eqref{q_ijkt_appendix} and the fact that Alice's answer is independent of the value of $\alpha$: $$q(i,t=0)=\frac{1}{\lambda_0^6+\lambda_1^6} p(a=\chi_i,(\alpha,\beta,\gamma)\in (X_0,Y_0 , Z_0))
=\frac{1}{\lambda_0^6+\lambda_1^6} \frac{P(X_0)}{P(X_1)} p(a=\chi_i,(\alpha,\beta,\gamma)\in (X_1,Y_0 , Z_0))=\frac{\lambda_0^6}{\lambda_0^6+\lambda_1^6} u_i^2,$$
where we used \eqref{PQi} and the fact that $P(X_0) =\lambda_0^2$ and $P(X_1) =\lambda_1^2$ for the last equality.
We can compute $q(i,t=1)$, \eqref{app:marg_q_jt_qubit} and  \eqref{app:marg_q_kt_qubit} in a similar manner.
\end{proof}

We are in position to conclude the proof. 
Indeed, the marginal conditions given above are incompatible when $u$ is chosen to be large enough, as given in Fig.~\ref{function}.
We have:

\begin{step}[Theorem 1.]
There exists no trilocal model reproducing $P_Q(a,b,c)$, whenever  $ u_{\mathrm{max}}^2(\lambda_0) < u^2 < 1$ (see Fig.~\ref{function}), where $u_{\mathrm{max}}(\lambda_0)$ is implicitly defined by the constraint~\eqref{ineq}.
\end{step}

\begin{figure}
\centering
\includegraphics[width=0.6\columnwidth]{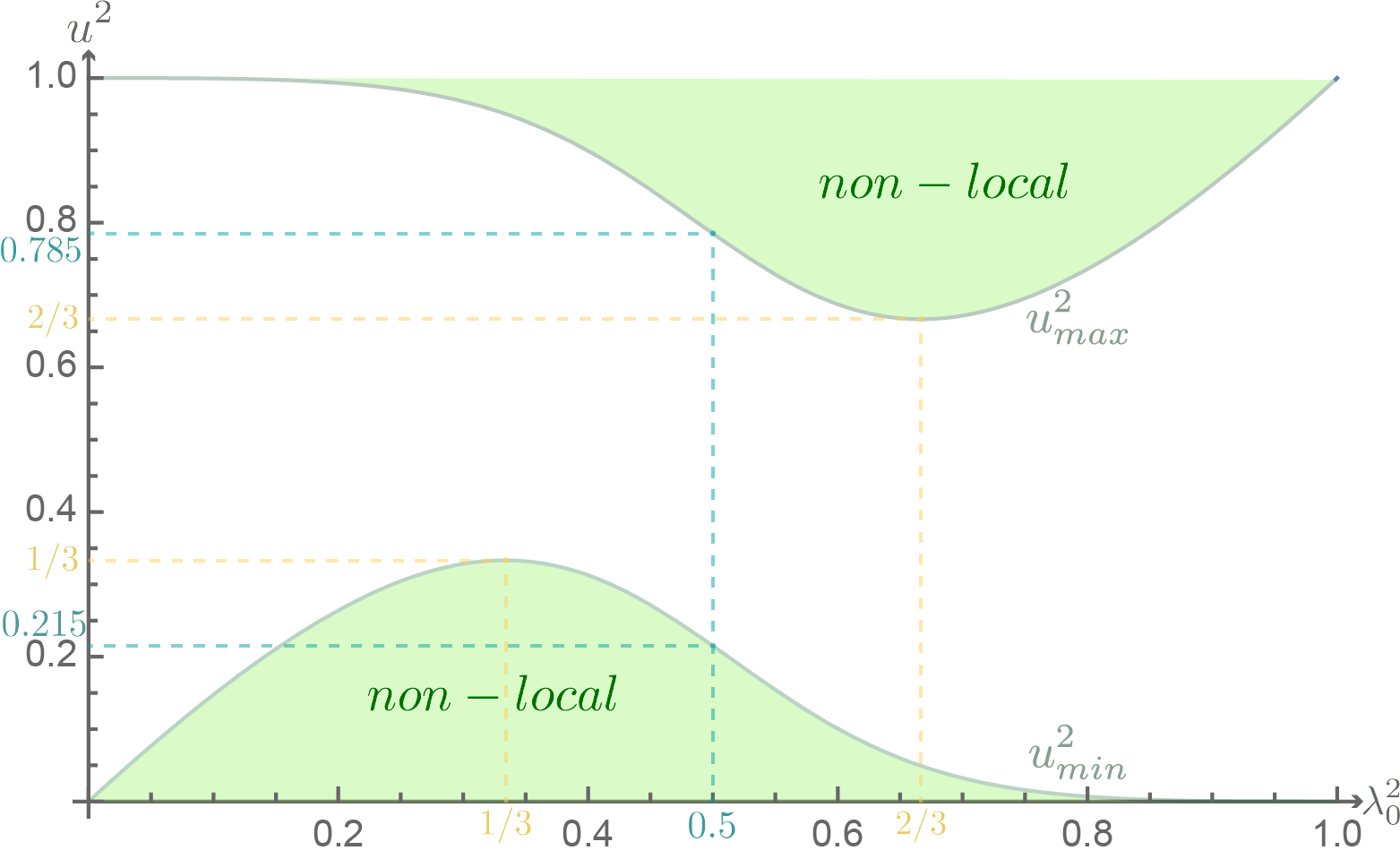}
\caption{Values of the squared measurement parameter $u^2$, as a function of the squared degree of entanglement $\lambda_0^2$ of the shared states, for which the quantum distribution $P_Q(a,b,c)$ can be proven to admit no trilocal model. Note that in this plot we also considered $u<v$.} \label{figure_elisa}
\label{function}
\end{figure}

\begin{proof}
Let us introduce a new probability distribution $\tilde{q}$, the symmetrization of $q$ over $i,j,k$:
\begin{equation}
\tilde{q}(i,j,k,t)=\frac{1}{6} \left(q(i,j,k,t)+q(j,k,i,t)+q(k,i,j,t)+q(i,k,j,t)+q(k,j,i,t)+q(j,i,k,t)\right).
\end{equation}
Clearly, $\tilde{q}$ still satisfies the same constraints as $q$ given in Step~\ref{step:appendix_qubit_triangle_2}. 
Let $\xi_{ijk}:=\tilde{q}(i,j,k,t=0)-\tilde{q}(i,j,k,t=1)$. We have
\begin{align}
\tilde{q}(i,j,k,t=0)=\frac{1}{2} \left( \tilde{q}(i,j,k) + \xi_{ijk} \right), ~~~~~~~~
\tilde{q}(i,j,k,t=1)=\frac{1}{2} \left( \tilde{q}(i,j,k) - \xi_{ijk} \right).
\end{align}
For simplicity, we write
$\xi_0:=\xi_{000}$, $\xi_{1}:=\xi_{100}=\xi_{010}=\xi_{001}$, $\xi_2:=\xi_{110}=\xi_{101}=\xi_{011}$ and $\xi_{3}:=\xi_{111}$.
Hence, with \eqref{app:marg_q_ijk_qubit}, we can write the marginals
\begin{align*}
\tilde{q}(k,t=0)&=\frac{1}{2} \sum_{i,j} \left(\frac{(\lambda_0^3 u_iu_ju_k + \lambda_1^3 v_iv_jv_k)^2 }{\lambda_0^6 + \lambda_1^6}  + \xi_{ijk} \right)=\frac{1}{2} \left( \frac{\lambda_0^6u_k^2+\lambda_1^6v_k^2}{\lambda_0^6+\lambda_1^6} + \sum_{ij} \xi_{ijk} \right)\\
\tilde{q}(k,t=1)&=\frac{1}{2} \sum_{i,j} \left(\frac{(\lambda_0^3 u_iu_ju_k + \lambda_1^3 v_iv_jv_k)^2 }{\lambda_0^6 + \lambda_1^6}  - \xi_{ijk} \right) = \frac{1}{2} \left( \frac{\lambda_0^6u_k^2+\lambda_1^6v_k^2}{\lambda_0^6+\lambda_1^6} - \sum_{ij} \xi_{ijk} \right) ,
\end{align*}
As $\sum_{ij} \xi_{ij0}=\xi_0+ 2 \xi_1+\xi_2$, $\sum_{ij} \xi_{ij1}=\xi_1+ 2 \xi_2+\xi_3$, we deduce
\begin{align}
\xi_0&=\frac{\lambda_0^6u^2-\lambda_1^6v^2}{\lambda_0^6+\lambda_1^6} - 2 \xi_1 - \xi_2=u^2-\frac{\lambda_1^6}{\lambda_0^6+\lambda_1^6} - 2 \xi_1 - \xi_2\\
\xi_3&=\frac{\lambda_0^6 v^2 - \lambda_1^6 u^2}{\lambda_0^6+\lambda_1^6} - \xi_1 - 2 \xi_2 = \frac{\lambda_0^6}{\lambda_0^6+\lambda_1^6} - u^2 -\xi_1 - 2 \xi_2 .
\end{align}
We have the positivity conditions
\begin{align}
0 \leq \tilde{q}(0,0,0,1)= \frac{(\lambda_0^3 u^3+\lambda_1^3 v^3)^2}{2(\lambda_0^6+\lambda_1^6)} - \frac{\xi_0}{2}, \quad
0 \leq \tilde{q}(1,1,1,0)=  \frac{(\lambda_0^3 v^3-\lambda_1^3 u^3)^2}{2(\lambda_0^6+\lambda_1^6)} + \frac{\xi_3}{2}
\end{align}
hence
\begin{align}
\xi_2 \geq u^2-\frac{\lambda_1^6 + (\lambda_0^3 u^3+\lambda_1^3 v^3)^2}{ (\lambda_0^6+\lambda_1^6)}  - 2 \xi_1, \quad
\xi_2 \leq \frac{\lambda_0^6 +(\lambda_0^3 v^3-\lambda_1^3 u^3)^2}{2(\lambda_0^6+\lambda_1^6)} - \frac{u^2}{2} - \frac{\xi_1}{2},
\end{align}
leading to
\begin{align}
\xi_1 \geq u^2 - \frac{2(\lambda_1^6 + (\lambda_0^3 u^3+\lambda_1^3 v^3)^2) + \lambda_0^6 +(\lambda_0^3 v^3-\lambda_1^3 u^3)^2}{3 (\lambda_0^6+\lambda_1^6)}.
\end{align}
Finally, we use the positivity condition
\begin{align}
0\leq \tilde{q}(0,0,1,1)= \frac{(\lambda_0^3 u^2 v - \lambda_1^3 u v^2)^2}{2(\lambda_0^6+\lambda_1^6)} - \frac{\xi_1}{2} 
\end{align}
to get 
\begin{align}
3(\lambda_0^3 u^2 v - \lambda_1^3 u v^2)^2- 3u^2 (\lambda_0^6+\lambda_1^6) + 2(\lambda_1^6 + (\lambda_0^3 u^3+\lambda_1^3 v^3)^2) + \lambda_0^6 +(\lambda_0^3 v^3-\lambda_1^3 u^3)^2 \geq 0.\label{ineq}
 \end{align}
This last inequality implicitly defines $u_\mathrm{max}(\lambda_0)$, which is plotted in Fig. \ref{function}.
\end{proof}

\setcounter{step}{0}
\section{Generalisation to qutrits}

In this section, we state and give the proof of a generalization of the previous example to the qutrit case. 

Now Alice, Bob and Charlie share two-qutrit maximally entangled states $ \ket{\psi} = \frac{1}{\sqrt{3}}(\ket{00} + \ket{11} + \ket{22})$. Each party performs the same joint entangled measurement, with nine outcomes denoted by $k \in \{ 
\tilde{0}, \tilde{1}, \tilde{2}, \chi_i^\uparrow, \chi_j^\downarrow \}$ with $i,j=0..2$. The corresponding eigenstates are given by
\begin{align*}
&\ket{\tilde{0}}:=\ket{00},~~  \ket{\tilde{1}}:=\ket{11},~~  \ket{\tilde{2}}:=\ket{22},\\
&\ket{\chi_i^\uparrow}=\eta_i^{01}\ket{01}+\eta_i^{02}\ket{02}+\eta_i^{12}\ket{12}, \\
&\ket{\chi_i^\downarrow}=\eta_i^{10}\ket{10}+\eta_i^{20}\ket{20}+\eta_i^{21}\ket{21},
\end{align*}
where the coefficients $\{\eta_i^{01},\eta_i^{02},\eta_i^{12}\}$ and $\{\eta_i^{10},\eta_i^{20},\eta_i^{21}\}$ are real and chosen such that the eigenstates $\{\ket{\chi_{i}^\uparrow}\}_{i=0..2}$ and $\{\ket{\chi_{i}^\downarrow}\}_{i=0..2}$ each form an orthonormal basis. This implies that the $\eta-$matrices are both orthonormal, meaning that their inverse matrices are their transposed matrices and therefore $\ket{01}=\sum_i\eta^{01}_i\ket{\chi_i^\uparrow}$ etc.

The global quantum state shared by the parties is given by
\begin{align*}
\ket{\psi_{ABC}}
=\frac{1}{3\sqrt{3}}&\Big(\ket{\tilde{0}}_A\ket{\tilde{0}}_B\ket{\tilde{0}}_C+\ket{\tilde{1}}_A\ket{\tilde{1}}_B\ket{\tilde{1}}_C+\ket{\tilde{2}}_A\ket{\tilde{2}}_B\ket{\tilde{2}}_C\\
&+\ket{\tilde{0}}_A\ket{01}_B\ket{10}_C+\ket{\tilde{0}}_A\ket{02}_B\ket{20}_C+\ket{\tilde{1}}_A\ket{12}_B\ket{21}_C\\
&+\ket{10}_A\ket{\tilde 0}_B\ket{01}_C+\ket{20}_A\ket{\tilde 0}_B\ket{02}_C+\ket{21}_A\ket{\tilde 1}_B\ket{12}_C\\
&+\ket{01}_A\ket{10}_B\ket{\tilde 0}_C+\ket{02}_A\ket{20}_B\ket{\tilde 0}_C+\ket{12}_A\ket{21}_B\ket{\tilde 1}_C\\
&+\ket{\tilde{1}}_A\ket{10}_B\ket{01}_C+\ket{\tilde{2}}_A\ket{20}_B\ket{02}_C+\ket{\tilde{2}}_A\ket{21}_B\ket{12}_C\\
&+\ket{01}_A\ket{\tilde 1}_B\ket{10}_C+\ket{02}_A\ket{\tilde 2}_B\ket{20}_C+\ket{12}_A\ket{\tilde 2}_B\ket{21}_C\\
&+\ket{10}_A\ket{01}_B\ket{\tilde 1}_C+\ket{20}_A\ket{02}_B\ket{\tilde 2}_C+\ket{21}_A\ket{12}_B\ket{\tilde 2}_C\\
&+\ket{01}_A\ket{12}_B\ket{20}_C+\ket{10}_A\ket{02}_B\ket{21}_C+\ket{02}_A\ket{21}_B\ket{10}_C\\
&+\ket{20}_A\ket{01}_B\ket{12}_C+\ket{12}_A\ket{20}_B\ket{01}_C+\ket{21}_A\ket{10}_B\ket{02}_C 
\Big)
\end{align*}
\begin{align*}
\ket{\psi_{ABC}}
=\frac{1}{3\sqrt{3}}&\Big(\ket{\tilde{0}}_A\ket{\tilde{0}}_B\ket{\tilde{0}}_C+\ket{\tilde{1}}_A\ket{\tilde{1}}_B\ket{\tilde{1}}_C+\ket{\tilde{2}}_A\ket{\tilde{2}}_B\ket{\tilde{2}}_C\\
&+\sum_{j,k}(\eta^{01}_j\eta^{10}_k+\eta^{02}_j\eta^{20}_k)\ket{\tilde{0}}_A\ket{\chi_j^\uparrow}_B\ket{\chi_k^\downarrow}_C+\eta^{12}_j\eta^{21}_k\ket{\tilde{1}}_A\ket{\chi_j^\uparrow}_B\ket{\chi_k^\downarrow}_C\\
&+\sum_{k,i}(\eta^{01}_k\eta^{10}_i+\eta^{02}_k\eta^{20}_i)\ket{\tilde{0}}_B\ket{\chi_k^\uparrow}_C\ket{\chi_i^\downarrow}_A+\eta^{12}_k\eta^{21}_i\ket{\tilde{1}}_B\ket{\chi_k^\uparrow}_C\ket{\chi_i^\downarrow}_A\\
&+\sum_{i,j}(\eta^{01}_i\eta^{10}_j+\eta^{02}_i\eta^{20}_j)\ket{\tilde{0}}_C\ket{\chi_i^\uparrow}_A\ket{\chi_j^\downarrow}_B+\eta^{12}_i\eta^{21}_j\ket{\tilde{1}}_C\ket{\chi_i^\uparrow}_A\ket{\chi_j^\downarrow}_B\\
&+\sum_{j,k}\eta^{10}_j\eta^{01}_k\ket{\tilde{1}}_A\ket{\chi_j^\downarrow}_B\ket{\chi_k^\uparrow}_C+(\eta^{20}_j\eta^{02}_k+\eta^{21}_j\eta^{12}_k)\ket{\tilde{2}}_A\ket{\chi_j^\downarrow}_B\ket{\chi_k^\uparrow}_C\\
&+\sum_{k,i}\eta^{10}_k\eta^{01}_i\ket{\tilde{1}}_B\ket{\chi_k^\downarrow}_C\ket{\chi_i^\uparrow}_A+(\eta^{20}_k\eta^{02}_i+\eta^{21}_k\eta^{12}_i)\ket{\tilde{2}}_B\ket{\chi_k^\downarrow}_C\ket{\chi_i^\uparrow}_A\\
&+\sum_{i,j}\eta^{10}_i\eta^{01}_j\ket{\tilde{1}}_C\ket{\chi_i^\downarrow}_A\ket{\chi_j^\uparrow}_B+(\eta^{20}_i\eta^{02}_j+\eta^{21}_i\eta^{12}_j)\ket{\tilde{2}}_C\ket{\chi_i^\downarrow}_A\ket{\chi_j^\uparrow}_B\\
&+\sum_{i,j,k}\eta^{01}_i\eta^{12}_j \eta_k^{20} \ket{\chi_i^\uparrow}_A \ket{\chi_j^\uparrow}_B \ket{\chi_k^\downarrow}_C+\eta^{10}_i\eta^{02}_j \eta_k^{21} \ket{\chi_i^\downarrow}_A \ket{\chi_j^\uparrow}_B \ket{\chi_k^\downarrow}_C\\
&+\sum_{i,j,k}\eta^{02}_i\eta^{21}_j \eta_k^{10} \ket{\chi_i^\uparrow}_A \ket{\chi_j^\downarrow}_B \ket{\chi_k^\downarrow}_C+\eta^{20}_i\eta^{01}_j \eta_k^{12} \ket{\chi_i^\downarrow}_A \ket{\chi_j^\uparrow}_B \ket{\chi_k^\uparrow}_C\\
&+\sum_{i,j,k}\eta^{12}_i\eta^{20}_j \eta_k^{01} \ket{\chi_i^\uparrow}_A \ket{\chi_j^\downarrow}_B \ket{\chi_k^\uparrow}_C+\eta^{21}_i\eta^{10}_j \eta_k^{02} \ket{\chi_i^\downarrow}_A \ket{\chi_j^\downarrow}_B \ket{\chi_k^\uparrow}_C\Big).
\end{align*}

From this last equation, we can determine a certain number of crucial properties of the distribution $P_Q(a,b,c)$. We observe that 
\begin{align}\label{constraint_trits_1}
\frac{1}{3^3}=P_Q(\tilde{0},\tilde{0},\tilde{0}) = \sqrt{P_Q(a=\tilde{0})P_Q(b=\tilde{0})P_Q(c=\tilde{0})} = \sqrt{\left(\frac{1}{3^2}\right)^3}.
\end{align}
Moreover, 
\begin{align}
\mathrm{for}~i\neq j, P_Q(a=\tilde{i},b=\tilde{j})&=0\label{constraint_trits_2}\\
p(b=\tilde{0},c\in\chi^\downarrow)=0\label{constraint_trits_3}\\
p(c\in\chi^\downarrow,a=\tilde{2})=0\label{constraint_trits_4}
\end{align}
We have other similar relations by permuting the parties. Finally, we have that 
\begin{align} 
P_Q(a=\chi_i^\uparrow,b=\tilde{1})=\frac{1}{3^3}\left(\eta_i^{01}\right)^2,~~P_Q(c=\tilde{1},a=\chi_i^\uparrow)=\frac{1}{3^3}\left(\eta_i^{12}\right)^2
\label{constraint_trits_5}\\
P_Q(a=\chi_i^\uparrow,b=\chi_j^\downarrow,c=\tilde{0})=\frac{1}{3^3}(\eta^{01}_i\eta^{10}_j+\eta^{02}_i\eta^{20}_j)^2. \label{constraint_trits_6}
\end{align}

We now can prove the following Theorem:
\begin{theorem}\label{theorem_qutrit_appendix}
For some choice of $\{\ket{\chi_i^\uparrow},\ket{\chi_i^\downarrow}\}$, the quantum distribution
$P_Q(a,b,c)$ is incompatible with any trilocal model.
\end{theorem}
~\\
Again, the proof of Theorem~\ref{theorem_qutrit_appendix} follows from two similar steps.
In Step~\ref{step:appendix_qutrit_1}, we consider the behavior for which the outputs $\{\chi_i^\uparrow\}$ and $\{\chi_i^\downarrow\}$ are grouped together into two outputs $\chi^\uparrow$ and $\chi^\downarrow$. We show that in that case, one only needs one trit of information about each of the local hidden variables to find the outputs.
Using this restriction and exploiting relation \eqref{constraint_trits_5}, \eqref{constraint_trits_6}, Step~\ref{step:appendix_qubit_triangle_2} shows that those restrictions can be exploited to compute marginals of a probability distribution grouping outputs and hidden variables. For a good choice of measurement basis, those marginals are incompatible. 

\begin{step}\label{step:appendix_qutrit_1}
We consider the coarse graining of the output set $\{\tilde{0},\tilde{1},\tilde{2},\chi^\uparrow,\chi^\downarrow\}$. The sources sets can be partitioned in three subsets
\begin{equation}\label{PartitionTrit}
X=X_0\amalg X_1\amalg X_2,~~~~~~~~Y=Y_0\amalg Y_1\amalg Y_2,~~~~~~~~Z=Z_0\amalg Z_1\amalg Z_2,
\end{equation}
(with $P(X_0)=\cdots=P(Z_2)=1/3$) such that the sets from which the local variables $ \alpha$, $\beta$ and $\gamma$ are taken determines all outputs.
More precisely, Alice answers 
\begin{enumerate}[label=(\roman*)]
\item $a = \tilde{i}$ if she receives $\beta\in Y_i$ and $ \gamma \in Z_i$, for $\tilde{i}\in\{\tilde{0},\tilde{1},\tilde{2}\}$, 
\item $a=\chi^\uparrow$ if she receives $\beta \in Y_j$ and $\gamma \in Z_k$ with $j<k$,
\item $a=\chi^\downarrow$ if she receives $\beta \in Y_j$ and $\gamma \in Z_k$ with $j>k$, 
\end{enumerate}
and similarly for Bob and Charlie (with a direct orientation of the cycle).
\end{step}

\begin{proof}

We first recall the Finner inequality.
It is shown in \cite{Renou} that any probability distribution admitting a classical model (it is also valid for a quantum or a ``boxworld'' model) satisfies the Finner inequality. More precisely, considering outputs for each party called $0$ here after, we have
\begin{equation}
P(000)\leq \sqrt{P_A(0)P_B(0)P_C(0)}.
\end{equation}
The equality condition is obtained iff there exist three subsets $X_0\subset X, Y_0\subset Y, Z_0\subset Z$ such that $A$ answers $0$ iff $\beta\in Y_0, \gamma\in Z_0$, $B$ answers $0$ iff $\gamma\in Z_0, \alpha\in X_0$ and $C$ answers $0$ iff $\alpha\in X_0,\beta\in Y_0$. In that case $P_A(0)=p(\beta\in Y_0)p(\gamma\in Z_0), P_B(0)=p(\gamma\in Z_0)p(\alpha\in X_0), P_C(0)=p(\alpha\in X_0)p(\beta\in Y_0)$.

In our case, \eqref{constraint_trits_1} shows equality in the Finner inequality for the outputs $(\tilde{0},\tilde{0},\tilde{0})$, hence we can introduce the corresponding sets $X_0,Y_0,Z_0$. Similarly, sets $X_1,Y_1,Z_1,X_2,Y_2,Z_2$ can be introduced.
Each of these sets have a probability $1/3$, and \eqref{constraint_trits_2} implies that they are disjoint.
Hence we have \eqref{PartitionTrit} and \textit{(i)}.
Moreover, if $\alpha\in X_0, \beta\in Y_1$, \eqref{constraint_trits_3} implies that $c(\alpha,\beta)\in \chi^\uparrow$.
If $\alpha\in X_1, \beta\in Y_2$, \eqref{constraint_trits_4} implies that $c(\alpha,\beta)\in \chi^\uparrow$.
Lastly, if $\alpha\in X_0, \beta\in Y_2$, \eqref{constraint_trits_3} (as well as \eqref{constraint_trits_4}) implies that $c(\alpha,\beta)\in \chi^\uparrow$.
Hence, by symmetry, we have \textit{(ii)} and \textit{(iii)}.
\end{proof}

\begin{step}\label{step_marginals_Trits_appendix}
Let us introduce 
\begin{align}
q(i,j,t)& :=p\big(a=\chi_i^\uparrow,b=\chi_j^\downarrow,c=\tilde{0},\gamma\in Z_t | \alpha\in X_0, \beta\in Y_0, \gamma\in Z_1\cup Z_2\big )\nonumber\\
&=\frac{3^3}{2} p\big(a=\chi_i^\uparrow,b=\chi_j^\downarrow,c=\tilde{0},\gamma\in Z_t) \label{q_ijkt_Trit_appendix}
\end{align}
$q(i,j,t)$ is a probability distribution. Moreover, the following marginal distributions of $q(i,j,t)$ satisfy:
\begin{align}
q(i,j)&= \sum_t q(i,j,t) = \frac{1}{2} (\eta^{01}_i\eta^{10}_j+\eta^{02}_i\eta^{20}_j)^2, \label{marg_q_ij_trit_appendix}\\
q(i,t=1)=&\frac{1}{2} (\eta^{01}_i)^2,~~~~ 
q(i,t=2)=\frac{1}{2} (\eta^{02}_i)^2,\label{marg_q_it_trit_appendix}\\
q(j,t=1)=&\frac{1}{2} (\eta^{10}_j)^2,~~~~ 
q(j,t=2)=\frac{1}{2} (\eta^{20}_j)^2.\label{marg_q_jt_trit_appendix}
\end{align}
\end{step}

\begin{proof}
Note that when $\alpha\in X_0, \beta\in Y_0, \gamma\in Z_1\cup Z_2$, we get $a\in\chi^\uparrow,b\in\chi^\downarrow,c=\tilde{0}$. Thus $q(i,j,t)$ is a probability distribution.

\eqref{marg_q_ij_trit_appendix} follows directly from \eqref{constraint_trits_6}. As $q(i,j,t)$ is a probability distribution, \eqref{marg_q_it_trit_appendix} can be deduced from Step~\ref{step:appendix_qutrit_1}, \eqref{marg_q_ij_trit_appendix} and the fact that Alice's answer is independent of the value of $\alpha$: 
\begin{align*}
q(i,t=1)&=\frac{3^3}{2} p(a=\chi_i^\uparrow,c=\tilde{0},\gamma\in Z_1))=\frac{3^3}{2} p(a=\chi_i^\uparrow,\alpha\in X_0,\beta\in Y_0,\gamma\in Z_1)\\&=\frac{3^3}{2} p(a=\chi_i^\uparrow,\alpha\in X_1,\beta\in Y_0,\gamma\in Z_1)=\frac{3^3}{2} p(a=\chi_i^\uparrow,b=\tilde{1})=\frac{1}{2}\left(\eta_i^{01}\right)^2
\end{align*}
The other marginal is simply given by $$q(i,t=2)=q(i)-q(i,t=1) =  \frac{1}{2}\left(\left(\eta_i^{01}\right)^2+\left(\eta_i^{02}\right)^2\right)-\frac{1}{2}\left(\eta_i^{01}\right)^2=\frac{1}{2}\left(\eta_i^{02}\right)^2,$$
where we used in the second equality that $\sum_j \left(\eta_j^{10}\right)^2 = \sum_j \left(\eta_j^{20}\right)^2=1$ and $\sum_j \eta_j^{10} \eta_j^{20} = 0$ due to orthonormality.
This proves (with symmetric arguments) Step~\ref{step_marginals_Trits_appendix}.
\end{proof}

We are in position to conclude the proof. Indeed, all we need now is to solve a Linear Programming problem. More precisely, the question is:\\
For all $\{\eta_i^{01},\eta_i^{02},\eta_i^{12}\}$ and $\{\eta_i^{10},\eta_i^{20},\eta_i^{21}\}$ such that the $\{\ket{\chi_i^\uparrow}\}$ and $\ket{\chi_i^\downarrow}\}$ form an orthonormal basis, does there exist a probability distribution $q(i,j,t)$ (linear conditions of positive coefficients summing to one) such that the marginal conditions (which are linear) \eqref{marg_q_ij_trit_appendix}, \eqref{marg_q_it_trit_appendix} and \eqref{marg_q_jt_trit_appendix} are satisfied?
In the following, we only exhibit one example for which this problem admits no solution, obtaining the following:

\begin{step}[Theorem 2.]
Consider 
\begin{align*}
\ket{\chi_0^\uparrow}&=\frac{1}{\sqrt{3}} \ket{01}+\frac{1}{\sqrt{2}}\ket{02}+\frac{1}{\sqrt{6}}\ket{12}, ~~
\ket{\chi_0^\downarrow}=\sqrt{\frac{2}{5}}\ket{10}+\sqrt{\frac{3}{5}}\ket{20},\\
\ket{\chi_1^\uparrow}&=\frac{1}{\sqrt{3}} \ket{01}-\frac{1}{\sqrt{2}}\ket{02}+\frac{1}{\sqrt{6}}\ket{12}, ~~
\ket{\chi_1^\downarrow}=\sqrt{\frac{3}{5}}\ket{10}-\sqrt{\frac{2}{5}}\ket{20}\\
\ket{\chi_2^\uparrow}&=\frac{1}{\sqrt{3}} \ket{01}-2\frac{1}{\sqrt{6}}\ket{12}, ~~~~~~~~~~~~~
\ket{\chi_2^\downarrow}=\ket{21}
\end{align*}
and $\ket{\chi_2^\uparrow}, \ket{\chi_2^\downarrow}$ chosen such that the $\{\ket{\chi_i^\uparrow}\}$ and $\ket{\chi_i^\downarrow}\}$ form an orthonormal basis.
There exists no trilocal model reproducing the corresponding $P_Q(a,b,c)$.
\end{step}
\begin{proof}
Following Step~\ref{step_marginals_Trits_appendix}, we find:
\begin{align*}
q(i=0,j=0)&=5/12,~~~~~~~~~~~~~~~~~~ q(i=0,j=1)=0,~~ &q(i=0,j=2)=0,\\
q(i=1,j=0)&=1/60,~~~~~~~~~~~~~~~~~~ q(i=1,j=1)=2/5,~~ &q(i=1,j=2)=0,\\
q(i=2,j=0)&=1/15,~~~~~~~~~~~~~~~~~~ q(i=2,j=1)=1/10,~~ &q(i=2,j=2)=0\\
\\
q(i=0,t=1)&=1/6,~~~~~~~~~~~~~~~~~~ q(i=1,t=1)=1/6,~~ &q(i=2,t=1)=1/6,\\
q(i=0,t=2)&=1/4,~~~~~~~~~~~~~~~~~~ q(i=1,t=2)=1/4,~~ &q(i=2,t=2)=0,\\
\\
q(j=0,t=1)&=1/5,~~~~~~~~~~~~~~~~~~ q(j=1,t=1)=3/10,~~&q(j=2,t=1)=0,\\
q(j=0,t=2)&=3/10,~~~~~~~~~~~~~~~~~~q(j=1,t=2)=1/5,~~ &q(j=2,t=2)=0.
\end{align*}
Let $M_t$, for $t = 1, 2$, be the $3\times 3$ matrix whose $(i, j)$-th entry equals $q(i, j, t)$ (for $0\leq i,j\leq 2$). Then the $i$-th row sum of $M_t$ equals $q(i, t)$, its $j$-th column sum is $q(j, t)$ and the $(i,j)$-th entry of $M_1 +M_2$ equals $q(i, j)$. Since $q(j = 2) = 0$, the last column of both $M_1,M_2$ is zero.
Also, since $q(i = 0, j = 1) = 0$, the $(0, 1)$ entry of both $M_1,M_2$ is zero. Finally since $q(i = 2, t = 2) = 0$ the last row of $M_2$ vanishes. Given these the other entries of $M_2$ will be determined uniquely; since there is only one non-zero entry in the second column of $M_2$, we must have $[M_2]_{(1,1)} = q(j = 1, t = 2) = 1/5$. By similar reasons the other two non-zero entries of $M_2$ can be determined. Moreover since $M_1 +M_2$ is known, $M_1$ can also be found. We obtain:
$$
M_1=\begin{pmatrix}
\frac{1}{6} & 0 & 0 \\
-\frac{1}{30} & \frac{1}{5} & 0 \\
\frac{1}{15} & \frac{1}{10} & 0
\end{pmatrix}, \qquad
M_2=\begin{pmatrix}
\frac{1}{4} & 0 & 0 \\
\frac{1}{20} & \frac{1}{5} & 0 \\
0 & 0 & 0
\end{pmatrix}.$$
We see that this implies that $q(i=1,j=0,k=1)=-1/30$ is negative, which is absurde, hence we have Theorem 2.
\end{proof}

\section{Generalisation to odd-cycle networks}\label{proof_odd}
\setcounter{step}{0}

We now give a generalization of Theorem 1 to any odd-cycle network. Here, we consider $N$ odd and parties $A_1, ..., A_N$ sharing maximally entangled qubit states in a cycle. The measurement performed by the parties are as in the main text, namely projective in the basis
\begin{align} 
\ket{\uparrow}&=\ket{01} &  \ket{\chi_0}=u_0 \ket{00} + v_0 \ket{11}  \nonumber \\
\ket{\downarrow}&=\ket{10} &  \ket{\chi_1}=u_1 \ket{00} + v_1 \ket{11} \label{app:BSM_trit}
\end{align}
with real numbers $u=u_0=-v_1$, $v=v_0=u_1$, $u_0^2+v_0^2=1$.

Let $P_Q(a_1,...,a_N)$ be the corresponding probability distribution.
It satisfies several properties. First, note that the number of parties that output $\chi$ (i.e. either $\chi_0$ or $\chi_1$) must be odd. Second, we observe that 
\begin{align}\label{app:P_QUpUp_trit}
P_Q(a_k=\uparrow,a_{k+1}=\uparrow) = P_Q(a_{k}=\downarrow,a_{k+1}=\downarrow)  =0.
\end{align}
Finally, we have that 
\begin{align} 
P_Q(\chi_i,\uparrow,\downarrow,\uparrow,...\uparrow,\downarrow)&=\frac{1}{2^N}u_i^2,\label{app:PQiN}\\
P_Q(a_1=\chi_{i_1}, a_2=\chi_{i_2},\dots, a_N=\chi_{i_N} )&=\frac{1}{2^N}\big(u_{i_1}u_{i_2}\dots u_{i_N} + v_{i_1}v_{i_2}\dots v_{i_N}\big)^2,
\label{app:PQi1...iN}
\end{align}
with other similar relations by permuting the parties. 
We have the following 
\begin{theorem}\label{theorem_qubit_odd}
For $u$ asymptotically close to $1$, there exists no $N$-cycle classical model reproducing $P_Q(a_1,...,a_N)$.
\end{theorem}
The proof, in two steps, is similar to the one of Theorem 1.  We first assume, by contradiction, that a classical $N$-local strategy exists.
In Step~\ref{step:appendix_qubit_N_1}, we consider the behavior for which the two outputs $\chi_0, \chi_1$ are grouped into a single output $\chi$. We show that in that case, one only needs one bit of information about each of the local hidden variables to find the outputs.
Using this restriction and exploiting relations \eqref{app:PQiN}, \eqref{app:PQi1...iN}, Step~\ref{step:appendix_qubit_N_2} shows that those restrictions can be exploited to compute marginals of a probability distribution grouping outputs and hidden variables. For a good choice of measurement basis, those marginals are incompatible. 

We note $\alpha^{k-1}\in X^{k-1}$ and $\alpha^{k}\in X^{k}$ the two sources placed just before and after party $A_k$, where the order is defined by a positive orientation of the cycle (where the indices are defined modulo $N$, see Fig.~\ref{fig:pentagon}). 

\begin{step}\label{step:appendix_qubit_N_1}
We consider the coarse graining of the output set $\{\uparrow,\downarrow,\chi=\{\chi_0,\chi_1\}\}$. The sources sets can be partitioned in two subsets
\begin{equation}
X^k=X^k_0 \amalg X^k_1,
\end{equation}
(with $P(X^k_0)=P(X^k_1)=1/2$), such that the sets from which the local variables $ \{\alpha^k\}$ are taken determines all outputs.
More precisely, $A_k$ answers 
\begin{enumerate}[label=(\roman*)]
\item $a_k = \uparrow$ if she receives $\alpha^{k-1}\in X^{k-1}_0$ and $\alpha^{k}\in X^{k}_1$, 
\item $a_k=\downarrow$ if she receives $\alpha^{k-1}\in X^{k-1}_1$ and $\alpha^{k}\in X^{k}_0$, 
\item $a_k = \chi$ otherwise.
\end{enumerate}
\end{step}

\begin{proof}
 It can be proven similarly to the proof of Step 1 of Theorem 1.
\end{proof}

\begin{step}\label{step:appendix_qubit_N_2}
Let us introduce
\begin{align}
q(i_1,i_2,\dots,i_N,t)& :=p\Big(a_1=\chi_{i_1},a_2=\chi_{i_2},\dots,a_N=\chi_{i_N}, (\forall 1 \leq k \leq N, \alpha^k \in X^k_t ) \nonumber\\ 
& ~~~~~~~~~~~~~~~~~~~~~~~~~~~~~~~~~~~~~~~~~~~~~~~|(\forall 1 \leq l \leq N, \alpha^l \in X^l_0) \text{ or } (\forall 1 \leq l \leq N, \alpha^l \in X^l_1) \Big)\\
 &= 2^{N-1} p\Big(a_1=\chi_{i_1},a_2=\chi_{i_2},\dots,a_N=\chi_{i_N}, (\forall 1 \leq k \leq N, \alpha^k \in X^k_t )\Big)
\end{align}
$q(i_1,i_2,\dots,i_N,t)$ is a probability distribution. Moreover, the following marginal distributions of $q(i_1,i_2,\dots,i_N,t)$ satisfy:
\begin{align}
q(i_1,i_2,\dots,i_N)&=\half\big(u_{i_1}u_{i_2}\dots u_{i_N} + v_{i_1}v_{i_2}\dots v_{i_N}\big)^2\label{marg_q_i1_iN} \\
q(i_k,t=0)&=\half u_{i_k}^2~~~~ q(i_k,t=1)=\half v_{i_k}^2\label{it_N}
\end{align}
for $1 \leq k \leq N$.
\end{step}

\begin{proof}
\eqref{app:PQi1...iN} direclty implies \eqref{marg_q_i1_iN}.

To prove (\ref{it_N}), we consider for simplicity the case in which $k=1$. We have
\begin{align*}
    q(i_1,t=0) &= \sum_{i_2,i_3,\dots,i_N \in \{0,1\}} q(i_1,i_2,\dots,i_N,t=0)
    =2^{N-1} p\big(a_1=\chi_{i_1}, \forall 1 \leq k \leq N: \alpha^k\in X^k_0\big)\\
    &=2^{N-1} p\big(a_1=\chi_{i_1}, \forall (k~\mathrm{odd}, l~\mathrm{even},~ 1 \leq k,l \leq N):\alpha^k\in X^k_0,\alpha^l\in X^l_1 \big)\\
    &=2^{N-1}p(a_1=\chi_{i_1}, \forall (k~\mathrm{odd}, l~\mathrm{even},~ 2 \leq k,l \leq N):a_k=\downarrow,a_l=\uparrow)\\
    &=2^{N-1} \frac{1}{2^N}u_{i_1}^2=\half u_{i_1}^2,
\end{align*}
where we used that $A_1$ is independent from the sources which are not connected to him in the third equality, and \eqref{app:PQiN} for the last. Fig.\ \ref{fig:pentagon} shows this step for the pentagon scenario.
\begin{figure}[h]
\centering
\includegraphics[scale=0.3]{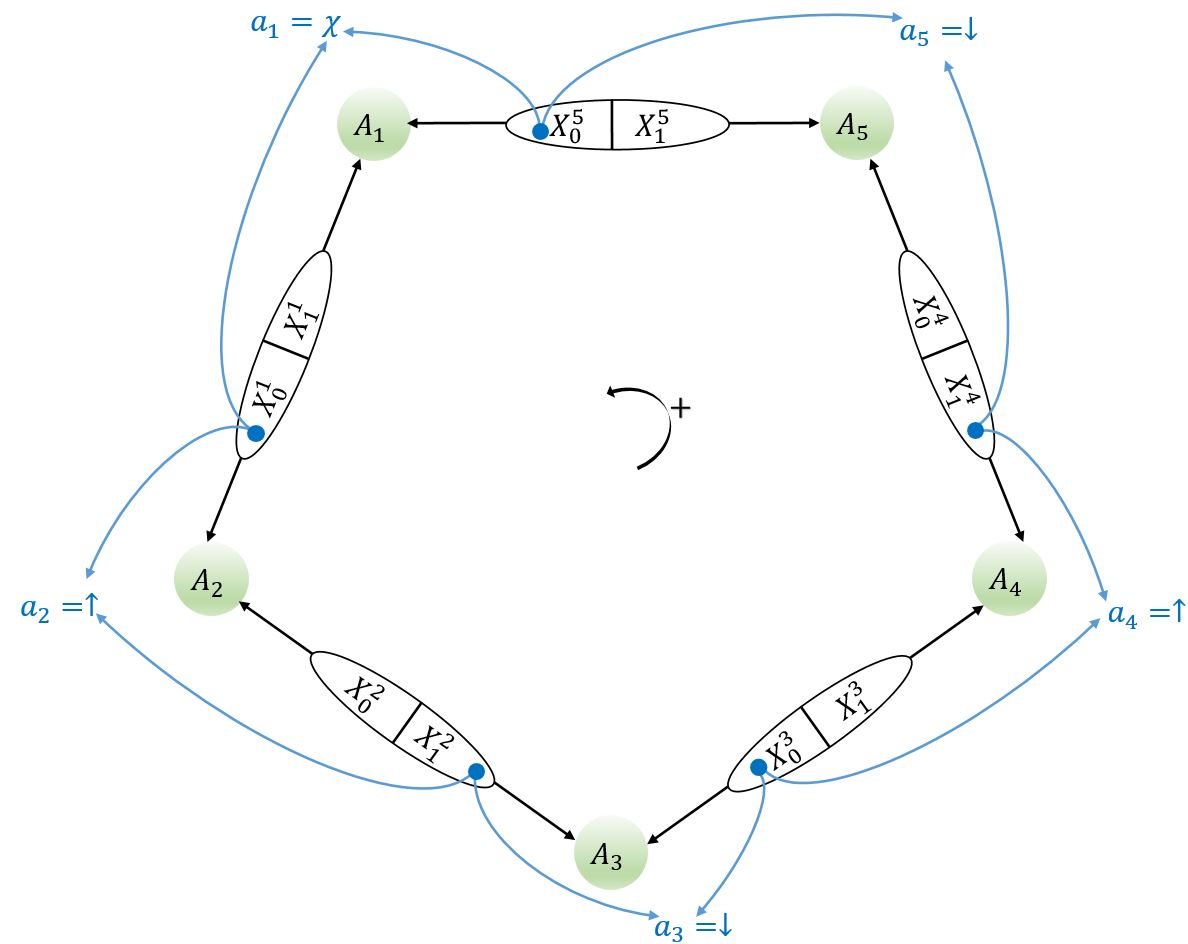}
\caption{Assigning $\alpha^k$s to $X^k_0$s and $X^k_1$s in the pentagon scenario $(N=5)$ to calculate  $q(i_1,t=0)$.}
\label{fig:pentagon}
\end{figure}
Similarly, we have  $q(i_k,t=0)=\half u_{i_k}^2$ and $q(i_k,t=1)=\half v_{i_k}^2$ for $1\leq k\leq N$.
\end{proof}
We now prove the theorem

\begin{step}[Theorem 3.]
For $u$ asymptotically close to $1$, there exists no $N$-cycle classical model reproducing $P_Q(a_1,...,a_N)$.
\end{step}
\begin{proof}
Let $\xi_{{i_1}{i_2}\dots{i_N}}$ be real numbers such that
\begin{equation}\label{q_x0_N}
    q({i_1}, {i_2}, \dots, {i_N} , t=0) = \half \Big( (u_{i_1}u_{i_2}\dots u_{i_N})^2 + u_{i_1}u_{i_2}\dots u_{i_N}v_{i_1}v_{i_2}\dots v_{i_N} + \xi_{{i_1}{i_2}\dots{i_N}}\Big)
\end{equation}

From $q({i_1}, {i_2}, \dots, {i_N})= q({i_1}, {i_2}, \dots, {i_N},t=0)+q({i_1}, {i_2}, \dots, {i_N},t=1)$ and Eq.(\ref{marg_q_i1_iN}) we have
\begin{equation}\label{q_x1_N}
    q({i_1}, {i_2}, \dots, {i_N} , t=1) = \half \Big( (v_{i_1}v_{i_2}\dots v_{i_N})^2 + u_{i_1}u_{i_2}\dots u_{i_N}v_{i_1}v_{i_2}\dots v_{i_N} - \xi_{{i_1}{i_2}\dots{i_N}}\Big)
\end{equation}

Using (\ref{it_N}) and the fact that $u_0^2+u_1^2 = v_0^2+v_1^2 = 1$ and $u_0v_0+u_1v_1=0$ we find that

\begin{equation}\label{sigma_x_N}
    \sum_{{i_2}\dots{i_N}} \xi_{{i_1}{i_2}\dots{i_N}}=0, \; \; \; \forall i_1\in \{0,1\}
\end{equation}
We define:
\begin{align*}
\tilde{q}({i_1}, {i_2}, \dots, {i_N} , t) = \frac{1}{N!} \sum_{\sigma\in \mathfrak{S}_N}q(i_{\sigma(1)}, {i_{\sigma(2)}}, \dots, {i_{\sigma(N)}} , t)\\
\tilde{\xi}({i_1}, {i_2}, \dots, {i_N} , t) = \frac{1}{N!} \sum_{\sigma\in \mathfrak{S}_N} \xi(i_{\sigma(1)}, {i_{\sigma(2)}}, \dots, {i_{\sigma(N)}}, t),
\end{align*}
where $\mathfrak{S}_N$ is the set of permutations of $1,...,N$.
Note that $\tilde{q}$ and $\tilde{\xi}$ are also probability distributions satisfying (\ref{q_x0_N}), (\ref{q_x1_N}), and (\ref{sigma_x_N}).
Let $S_d$ be the set of all binary strings of length $N$ with $d$ number of one ($|S_d|=\binom{N}{d}$).
By definition of $\tilde{q}$ we have $\tilde{q}(r,t)=\tilde{q}(s,t)$ for $r,s\in S_d$. We call it $Q_d(t)$. In a similar manner, we define $\xi_d(t)=\tilde{\xi}(s,t)$ for $s\in S_d$. From (\ref{q_x0_N}) and (\ref{q_x1_N}), for all $ 0\leq d \leq N~$we have:
\begin{align*}
Q_d(0)=\half (u^{2(N-d)}v^{2d}+(-1)^d u^N v^N+\xi_d)\\
Q_d(1)=\half (v^{2(N-d)}u^{2d}+(-1)^d u^N v^N-\xi_d)
\end{align*}
As we assume these probabilities come from a local hidden variable model, there should exist $\xi_0,\dots,\xi_N$ s.t. $Q_d(t) \geq 0, ~ \forall \; 0\leq d \leq N,\; t\in \{0,1\}$. Hence, we have:
\begin{equation*}
    v^{2(N-d)}u^{2d}+(-1)^d u^N v^N \geq \xi_d \geq -u^{2(N-d)}v^{2d}-(-1)^d u^N v^N
\end{equation*}
We consider $v=\epsilon$ and  $u=\sqrt{1-\epsilon^2}$ and an asymptotically small $\epsilon > 0 $. The Nth-order approximation (in terms of $\epsilon$) of these bounds implies
\begin{align} 
    \text{for } 0 \leq d \leq (N-1)/2 &: \text{  } 
         (-1)^d \epsilon^N \geq \xi_d,  \label{epsilonBound_N1} \\
    \text{for }  (N+1)/2 \leq d \leq N&: \text{  } 
     \xi_d \geq (-1)^{d+1} \epsilon^N, \label{epsilonBound_N2}
\end{align}
as $N$ is odd. Let $M=(N-1)/2$.
From (\ref{sigma_x_N}) we have
\begin{align*}
    \Gamma_0 &= \binom{2M}{0}\xi_0+\binom{2M}{1}\xi_1+\binom{2M}{2}\xi_2+\dots+\binom{2M}{2M}\xi_{2M}  = 0,\\
\Gamma_1 &= \binom{2M}{0}\xi_1+\binom{2M}{1}\xi_2+\binom{2M}{2}\xi_3+\dots+\binom{2M}{2M}\xi_{2M+1}  = 0.
\end{align*}
We now consider linear combination of these two last equations. Consider
\begin{align}\label{sigma_xixj}
    0 = \beta \Gamma_1 - \gamma \Gamma_0 = 
    -\sum_{i=0}^{M}\Big({\binom{2M}{i}\gamma-\binom{2M}{i-1}\beta}\Big)\xi_i+\sum_{j=M+1}^{2M+1}\Big({\binom{2M}{j-1}\beta-\binom{2M}{j}\gamma}\Big)\xi_j,
\end{align}
where we take the convention $\binom{2M}{2M+1} = \binom{2M}{-1}= 0$.
Consider now some $\beta>0,\gamma>0$ such that 
\begin{align}
\binom{2M}{M}\gamma-\binom{2M}{M-1}\beta \geq 0~~\mathrm{and}~~
\binom{2M}{M}\beta-\binom{2M}{M-1}\gamma \geq 0.\label{positivity_const}
\end{align}
This clearly exists (e.g. for $\beta,\gamma$ almost equal as $\binom{2M}{M}>\binom{2M}{M-1}$), and implies for $0 \leq i \leq M$ and $M+1 \leq j \leq 2M+1$,
\begin{align*}
\binom{2M}{i}\gamma-\binom{2M}{i-1}\beta \geq 0~~\mathrm{and}~~
\binom{2M}{j-1}\beta-\binom{2M}{j}\gamma \geq 0.
\end{align*}

Now, by using the inequalities (\ref{epsilonBound_N1}) and (\ref{epsilonBound_N2}) in (\ref{sigma_xixj}) we have:
\begin{align*}
    & 0 \geq
    \left[ -\sum_{i=0}^{M}(-1)^{i}\left({\binom{2M}{i}\gamma-\binom{2M}{i-1}\beta}\right)-\sum_{j=M+1}^{2M+1}(-1)^{j}\left({\binom{2M}{j-1}\beta-\binom{2M}{j}\gamma}\right) \right]\epsilon^{2M+1}
    +O(\epsilon^{2M+2})\\
    \geq & \left[\gamma\left(\sum_{j=M+1}^{2M+1}(-1)^{j}\binom{2M}{j}-\sum_{i=0}^{M}{(-1)^{i}\binom{2M}{i}}\right)+\beta\left(\sum_{i=0}^{M}{(-1)^{i}\binom{2M}{i-1}}-\sum_{j=M+1}^{2M+1}(-1)^{j}\binom{2M}{j-1}\right)\right]\epsilon^{2M+1}
    +O(\epsilon^{2M+2})
\end{align*}
Hence, we have
\begin{equation}\label{inequality0_epsilon}
    0 \geq  (\gamma - \beta) A \epsilon^{N} + O(\epsilon^{N+1}),
\end{equation}
where
\begin{align*}
    A = -\sum_{i=0}^{M}{(-1)^{i}\binom{2M}{i}}+\sum_{i=M+1}^{2M}(-1)^{i}\binom{2M}{i} 
    =(-1)^{M+1} \binom{2M}{M}.
\end{align*}

Hence for $N\equiv 1 \text{    (mod 4)}$, $A < 0$, and for $N\equiv 3 \text{    (mod 4)}$, $A > 0$.
Considering $\beta > \gamma$ for $N\equiv 1 \text{    (mod 4)}$, and $\beta < \gamma$ for $N\equiv 3 \text{    (mod 4)}$, such that \eqref{positivity_const} is satisfied (which always exists), we have $(\gamma - \beta) A > 0$,  \eqref{inequality0_epsilon} does not hold for sufficiently small $\epsilon > 0 $. Therefore we obtain a contradiction, which proves Theorem~\ref{theorem_qubit_odd}.
\end{proof}

\section{Trilocal model}

Here, we construct a trilocal model for the quantum distribution $P_Q(a,b,c)$ considering maximally entangled states for the sources (i.e. $\lambda_0^2=\lambda_1^2=1/2$) and measurements given by $u^2=u_{\mathrm{max}}^2 = \frac{-3+(9+6\sqrt{2})^{2/3}}{2(9+6\sqrt{3})^{1/3}}$. That is, the measurement parameter is chosen as to saturate the bound given in Theorem 1. 

The constraint \eqref{app:marg_q_ijk_qubit}, \eqref{app:marg_q_it_qubit}, \eqref{app:marg_q_jt_qubit}, \eqref{app:marg_q_kt_qubit} over $q(i,j,k,t)$ found in Step~\ref{step:appendix_qubit_triangle_2} write:
\begin{align}
q(i,j,k&) = \frac{1}{2}( u_i u_j u_k +  v_i v_j v_k )^2,\label{app:eq_model1}\\
q(i,t=0)& =\frac{1}{2} u_i^2, ~~~~q(i,t=1) =\frac{1}{2} v_i^2,\label{app:eq_model2}\\
q(j,t=0)& =\frac{1}{2} u_j^2, ~~~~q(j,t=1) =\frac{1}{2} v_j^2,\label{app:eq_model3}\\
q(k,t=0)& =\frac{1}{2} u_k^2, ~~~~q(k,t=1) =\frac{1}{2} v_k^2.\label{app:eq_model4}
\end{align}
For a trilocal strategy satisfying Step~\ref{step:appendix_qubit_triangle_1}, we write $P_0=\{p_0(i,j,k)\}$ (resp. $P_1=\{p_1(i,j,k)\}$) the output distributions obtained when the hidden variables are uniformly taken in the sets $X_0,Y_0,Z_0$ (resp. $X_1,Y_1,Z_1$). 
Of course, when a trilocal strategy satisfying Step~\ref{step:appendix_qubit_triangle_1} exists and gives $P_Q$, the distribution $q(i,j,k,t)$ given by $q(i,j,k,t)=1/2~p_t(i,j,k)$ satisfies \eqref{app:eq_model1}, \eqref{app:eq_model2}, \eqref{app:eq_model3}, \eqref{app:eq_model4}.

Conversely, assume we have $P_0=\{p_0(i,j,k)\}$ and $P_1=\{p_1(i,j,k)\}$ such that $q(i,j,k,t)=1/2~p_t(i,j,k)$ satisfies those four equations. Then it can be seen easily (by computing the output distribution) that the strategy where each party outputs as follows:
\begin{enumerate}
\item $\uparrow$, $\downarrow$ according to Step~\ref{step:appendix_qubit_triangle_1}.
\item according to the strategy for $P_0$ when they receive local variables in sets with same subscript $0$,
\item according to the strategy for $P_1$ when they receive local variables in sets with same subscript $1$,
\end{enumerate}
is a valid strategy which gives $P_Q$.

We now show that this is feasible for the case $u^2=u_{\mathrm{max}}^2 $. There, it is sufficient to consider symmetric $q$ (in our notations, $q=\tilde{q}$).
The constraints derived on $q$ in Appendix~\ref{mainresult} indicate that $q(0,0,0,1)=q(0,0,1,1)=q(1,1,1,0)=0$. Hence we introduce the following symmetric strategies for $P_t$, $t=0,1$:
\begin{itemize}
\item For $t=0$ the local variables take value in $\{0,1,2\}$, with distribution $\kappa_0,\kappa_1,\kappa_2$ where $\kappa_1=(1-u_{\mathrm{max}}^2)/(2\kappa_0)$ and $\kappa_2=1-\kappa_0-\kappa_1$. A party outputs $\chi_1$ if and only if he receives the trits $0,1$ or $1,0$.
\item For $t=1$ the local variables take value in $\{0,1\}$, with distribution $\tau_0,\tau_1$ with $\tau_1=1-\tau_0$. A party outputs $\chi_0$ if and only if he receives the trits $0,1$ (i.e.~0 from the left and 1 from the right).
\end{itemize}
Using the symbolic computation software Mathematica, we see that there exist $\kappa_0, \tau_0$ such that the corresponding $P_0=\{p_0(i,j,k)\}$, $P_1=\{p_1(i,j,k)\}$ are such that with $q(i,j,k,t)=1/2~p_t(i,j,k)$, $q$ satisfy the four equations \eqref{app:eq_model1}, \eqref{app:eq_model2}, \eqref{app:eq_model3}, \eqref{app:eq_model4}. For $t=0$ we where not able to analytically simplify the expressions and our claim rely on numerical approximation. For $t=1$, our claim can be verified analytically.

\end{document}